\documentclass[11pt]{article}
\usepackage[colorlinks]{hyperref}
\usepackage{amsmath}
\usepackage{amsthm}
\usepackage{amssymb}
\usepackage{graphicx}
\usepackage{algorithm}
\usepackage{algpseudocode}
\makeatletter
\renewcommand{\theHALG@line}{\thealgorithm.\arabic{ALG@line}}
\makeatother
\usepackage{multicol}
\usepackage{multirow}
\usepackage{color}
\usepackage{tikz}
\usepackage{xifthen}
\usetikzlibrary{decorations.pathreplacing,arrows.meta}

\usepackage[letterpaper,margin=1in,bottom=1in]{geometry}
\usepackage[capitalize]{cleveref}
\usepackage[utf8]{inputenc}
\usepackage[english]{babel}
\usepackage{mathtools}

\usepackage{authblk}
\usepackage{interval}
\intervalconfig{soft open fences}

\newtheorem{theorem}{Theorem}[section]

\newtheorem{lemma}[theorem]{Lemma}
\newtheorem{corollary}[theorem]{Corollary}
\newtheorem{proposition}[theorem]{Proposition}

\newtheorem{example}{Example}

\AddToHook{env/axiom/begin}{\crefalias{theorem}{axiom}}
\AddToHook{env/lemma/begin}{\crefalias{theorem}{lemma}}
\AddToHook{env/corollary/begin}{\crefalias{theorem}{corollary}}
\AddToHook{env/proposition/begin}{\crefalias{theorem}{proposition}}
\AddToHook{env/fact/begin}{\crefalias{theorem}{fact}}
\AddToHook{env/claim/begin}{\crefalias{theorem}{claim}}
\AddToHook{env/definition/begin}{\crefalias{theorem}{definition}}
\AddToHook{env/conjecture/begin}{\crefalias{theorem}{conjecture}}

\usepackage{tabularx}
\usepackage{booktabs}
\usepackage{threeparttable}
\usepackage{adjustbox}

\usepackage{microtype}
\usepackage{needspace}
\usepackage{xurl}
\DeclareMathOperator{\poly}{poly}

\newcommand{\F}{\mathcal F}
\newcommand{\XSAT}{\mathsf{XSAT}}
\newcommand{\satcount}[1]{\#\mathrm{SAT}(#1)}
\newcommand{\rate}[1]{\frac{#1\log\log(#1+16)}{\sqrt{\log(#1+16)}}}
\newcommand{\ind}{\mathbf 1}

\newcommand{\Nzero}{\mathbb N_0}
\newcommand{\templ}{\mathcal D}
\newcommand{\terms}{\mathcal T}

\hypersetup{
  hypertexnames=false,
  pdftitle={Subexponential-Time Exact-One Hitting Set},
  pdfauthor={Shiteng Chen, Tao Jiang, Yijie Zhu},
  pdfsubject={A linear-variable two-query exact-counting reduction},
  pdfkeywords={Exact-One Hitting Set, exact counting, exact satisfiability, subexponential algorithms, counting Turing reductions}
}

\begin{document}
\title{Subexponential-Time Exact-One Hitting Set}
\author[1,2,3]{Shiteng Chen}
\author[1,2,3]{Tao Jiang}
\author[1,2,3]{Yijie Zhu}
\affil[1]{Key Laboratory of System Software (Chinese Academy of Sciences), Beijing, China}
\affil[2]{Institute of Software, Chinese Academy of Sciences, Beijing, China}
\affil[3]{University of Chinese Academy of Sciences, Beijing, China}
\date{}
\maketitle
\begingroup
\let\small\normalsize
\begin{abstract}
Exact-One Hitting Set is one of Karp's original 21 NP-complete problems. Given an explicitly listed family of subsets of an $r$-element ground set, it asks whether some selection meets every input set exactly once. Using the fixed-width exact-counting theorem of OpenAI (2026), we obtain deterministic polynomial-space exact counting in time
\[
 \poly(L+1)\,2^{O\!\left(\rate{r}\right)}=2^{o(r)}\poly(L+1),
\]
where $L$ is the input length, with no restriction on the number or widths of the sets.

We develop compact representations behind this consequence. A path realization of two-query counting cancellation, combined with existing row selection, produces two ordinary $3$-CNF formulas with at most $2r$ variables and $O(r^2)$ clauses each. The same variable budget handles signed equations with fixed bounded demands and positive integer coefficients, using a larger fixed clause width. We also prove that every explicitly listed Boolean relation of VC dimension at most one can be represented, on its original variables, by binary clauses and signed exact-one equations. For arbitrarily many such local constraints with unrestricted overlaps, this gives two $3$-CNF queries of at most $2r$ variables each.

A bounded-term version of the path identity further handles fixed finite/cofinite local degree rules on arbitrary graphs. Functional threshold bits give linear-variable counting reductions, including size distributions; induced matching and perfect matching cuts provide compact examples. The path construction is a variable-eliminated realization of marker--parity cancellation. The comparison isolates the explicit variable budgets and structural representations from the algorithmic consequences already implied by existing counting tools.

\end{abstract}
\endgroup
\noindent\textbf{Keywords.} Exact-One Hitting Set, exact counting, constraint representations, VC dimension, local degree constraints, subexponential algorithms.
\clearpage
\setcounter{tocdepth}{1}
\tableofcontents
\clearpage
\section{Introduction}\label{sec:intro}

\emph{Exact-One Hitting Set} is one of Karp's original 21 NP-complete problems, listed as \emph{Hitting Set} in his 1972 paper~\cite[p.~95, Problem~15]{Karp72}. Given a ground set $U$ and an explicitly listed family $\F$ of subsets of $U$, it asks whether there exists $W\subseteq U$ such that
\begin{equation}\label{eq:exactone}
 |W\cap S|=1\qquad\text{for every }S\in\F.
\end{equation}
The exact-one requirement belongs to Karp's original formulation. There is no cardinality budget on $W$: coverage must be reconciled with exclusion, because choosing one element of a set forbids choosing any other element of that set. We measure complexity in $r:=|U|$, not in the number of constraints or incidences. The explicit input length $L$ includes all variable declarations. We compute the stronger exact count
\begin{equation}\label{eq:count}
 Z(U,\F):=\bigl|\{W\subseteq U:|W\cap S|=1\text{ for every }S\in\F\}\bigr|;
\end{equation}
positivity decides feasibility.

The fixed-width exact-counting algorithm of OpenAI~\cite[Theorem~1.1]{OAI26} makes a linear target-variable budget particularly useful. Its general-density bound gives the following algorithmic consequence of the reduction developed below.

\begin{theorem}[Exact-One Hitting Set]\label{thm:main}
For an explicitly represented Exact-One Hitting Set instance on $r$ elements and of input length $L$, one can compute $Z(U,\F)$ deterministically in polynomial space and time
\begin{equation}\label{eq:main-time}
 \poly(L+1)\,2^{O\!\left(\rate{r}\right)}=2^{o(r)}\poly(L+1).
\end{equation}
The number and widths of the input sets are unrestricted.
\end{theorem}

The time bound alone is not our novelty claim. A public marker--parity construction~\cite{Ono26} eliminates $h$ wide clauses on $n$ variables by subtracting two ordinary model counts with $n+2h-1$ variables. Combining that construction with the row-selection argument of Jansen and Pieterse~\cite{JP16,JP19} already yields the same asymptotic rate for Exact-One Hitting Set. We give a path realization with $n+h-1$ variables and use its explicit syntax to develop representation results for further constraint classes. Section~\ref{subsec:marker-relation} makes the relationship to marker--parity cancellation precise; a constant-factor improvement in variables is not claimed to change the order of the displayed time exponent.

The encoding issue is nevertheless substantive. Row selection retains at most $r+1$ original exact-one equations with exactly the same Boolean solutions, but their total width may still be $\Theta(r^2)$. A separate functional circuit for every retained constraint can introduce quadratically many variables. The alternative is to retain pairwise exclusions on the original variables and represent the remaining coverage conditions jointly. Two boundary conditions on a Boolean path give identical contributions to any assignment missing a condition; their difference retains each valid assignment once. Here $\satcount{G}$ counts assignments to all declared variables, and $3$-CNF means width at most three.

\begin{theorem}[Linear-variable exact-one reduction]\label{thm:reduction}
In polynomial time, an Exact-One Hitting Set instance on $r$ declared elements can either be solved directly when its family is empty or contains an empty set, or represented by two ordinary unweighted $3$-CNFs $Q_0,Q_1$ satisfying
\begin{equation}\label{eq:main-identity}
 Z(U,\F)=\satcount{Q_0}-\satcount{Q_1}.
\end{equation}
Each query has at most $2r$ variables and $O(r^2)$ clauses. The queries are nonadaptive and reconstruction uses polynomial-time integer arithmetic.
\end{theorem}

The general identity accepts a bounded-width CNF together with $h$ conditions, each an explicit disjunction of constant-length conjunctions. It uses $n+h-1$ variables: the number of alternative witnesses affects clauses, not variables. This also clarifies what a representation theorem must provide---few simultaneous conditions and short local witnesses---and leads to two extensions closely tied to exact satisfiability.

First, an equation may demand a fixed number of true literals rather than exactly one. More generally, allow positive binary-encoded integer coefficients and signed literals, with a fixed bound $d$ on the normalized demand of each row. After removing constants and choosing the smaller complementary demand, row selection again leaves at most $r+1$ rows. Bounded-width exclusions prevent excess, while a disjunction of terms of length at most $d$ prevents deficit. Theorem~\ref{thm:fixed-demand} gives two queries with at most $2r$ variables and width $\max\{3,d+2\}$. The demands can differ from row to row; the bound $d$, not an individual row value, is fixed. This extends the generalized exact-satisfiability formulation studied by Hoi and Stephan~\cite{HS21} to positive integer coefficients and exact counting.

Second, exact-one structure can be recovered from an allowed table rather than supplied as an equation. A Boolean relation has \emph{VC dimension at most one} when no pair of coordinates realizes all four Boolean patterns. We prove that every nonempty explicitly listed relation with this property is represented, on its original variables, by unary or binary clauses together with signed exact-one equations. The construction first excludes missing one- and two-coordinate patterns, then separates each remaining unwanted pattern by an exact-one equation. This yields the following consequence.

\begin{theorem}[Locally VC-dimension-one constraints]\label{thm:vc-intro}
Consider $r$ declared Boolean variables and any explicitly listed family of allowed tables on subsets of those variables. If each table has VC dimension at most one, the number of global satisfying assignments is recoverable in polynomial time from at most two ordinary $3$-CNF counts, each on at most $2r$ variables and with $O(r^2)$ clauses. It is computable within the time and space bounds of Theorem~\ref{thm:main}. The number of tables, their arities, and their overlaps are unrestricted.
\end{theorem}

The dimension assumption is local: the global solution space need not have small VC dimension or only polynomially many assignments. The representation is also more specific than arbitrary affine separation. Structure theorems for dimension-one classes~\cite{BenDavid15} and the interpolation-degree bound of Moran and Rashtchian~\cite{MR20} already explain their low-dimensional algebraic behavior. Our proof provides the particular binary-clause and signed-exact-one syntax needed to preserve the $2r$ query budget, with an explicit polynomial-time construction. No general priority claim about all equivalent representations is required for the reduction.

Fixed local graph rules supply another source of linearly many conditions. For fixed finite or cofinite sets $\sigma,\rho\subseteq\Nzero$, a selected set $S$ must satisfy $|N(u)\cap S|\in\sigma$ when $u\in S$ and $|N(u)\cap S|\in\rho$ otherwise. Fixed-class degree-constraint matrices impose an allowed neighbor-count set for each pair of classes. These are established frameworks~\cite{BXTV13,FMMNSSW25}. A fixed cutoff $D$ lets us encode the truth of ``at least $j$ selected neighbors'' for $1\le j\le D$ using $D$ functional threshold bits per vertex. Section~\ref{sec:local} proves deterministic polynomial-space exact counting in $\poly_{\mathrm{rule}}(L+1)2^{O_{\mathrm{rule}}(\eta(v))}$ time on arbitrary $v$-vertex graphs, where $\eta(v)=v\log\log(v+16)/\sqrt{\log(v+16)}$. Input size bounds are allowed; the local rules and class count remain fixed. Induced matching and perfect matching cuts illustrate the construction, while upper domination illustrates the separate role of nonunique decision certificates.

The central organization is thus representational: a compact realization of counting cancellation, bounded-demand equations, a syntax theorem for local dimension-one tables, and functional local-degree encodings. Appendix~\ref{app:weights} records a linear-size arithmetic tool for polynomially bounded integer weights and a one-query implementation of bounded-demand equations. It is not needed by the main two-query proofs. Additional graph encodings are collected in Appendix~\ref{app:graphs} rather than treated as separate headline contributions.

Exact-One Hitting Set is monotone Exact Satisfiability; allowing negative literals gives signed $\XSAT$. Incidence transposition gives Exact Cover~\cite{Knuth00}, with our parameter corresponding to the number of candidate sets, not universe size. Positive one-in-three constraints are NP-complete~\cite{Schaefer78}, and the corresponding counting problem is $\#\mathrm P$-complete~\cite{CH96}. Earlier exact algorithms distinguish decision and counting~\cite{DJB04,BMS05}; representative bounds are $O^*(1.1674^r)$ for XSAT decision~\cite{Hoi20}, $O^*(1.1995^r)$ for unrestricted-width counting~\cite{ZSW14}, and $O^*(1.1120^r)$ for width-three counting~\cite{HJS20}. Here $O^*$ suppresses input-polynomial factors. Clause-count algorithms~\cite{BH08} and structurally restricted subexponential results~\cite{PSSW14,Schuh18} address different parameter guarantees.

The imported counter and row selection retain their original attribution. Existing graph models include vertex-based induced matching~\cite{AET18}, regular induced subgraphs~\cite{GRS12}, matching cuts~\cite{LT22,MM23}, private neighbors~\cite{BHN23}, and the kernel CNF~\cite{WD12}. Recent ordinary-domination consequences~\cite{Anth26}, the public marker construction~\cite{Ono26}, and explicit minimal-transversal and upper-domination results~\cite{PI26} preclude treating every listed application as a first subexponential algorithm. We instead state the exact representations, query budgets, and count-preservation properties proved here.

Section~\ref{sec:overview} explains the mechanism. Sections~\ref{sec:prelim}--\ref{sec:identity} give the shared tools. Section~\ref{sec:exactone} treats exact equations, Section~\ref{sec:vc} proves the local-table representation, and Section~\ref{sec:local} treats local graph rules. Section~\ref{sec:graphs} presents representative applications, followed by scope and appendices.

\section{Technical overview}\label{sec:overview}

The reduction trades unique auxiliary extensions for controlled cancellation. This is a compact path realization of marker--parity cancellation~\cite{Ono26}; the formal variable elimination is given in Section~\ref{subsec:marker-relation}. To see why this helps, begin with an exact-one system after row selection. Pairwise clauses express that each retained set contains at most one chosen element, using no new variables. At most $r+1$ coverage clauses remain. A separate circuit for each clause would pay for its width. Instead, we allow a noncovering assignment to have extensions in both queries and arrange for their contributions to be identical.

Write the remaining formula as $B(x)\wedge C_1(x)\wedge\cdots\wedge C_h(x)$, where $B$ already has width at most three. Associate $C_j$ with edge $j$ of a Boolean path $z_0,\ldots,z_h$. A true literal of $C_j$ forces the two endpoints of that edge to be equal; false literals impose no restriction. The implication
\[
 \ell\ \Longrightarrow\ (z_{j-1}=z_j)
\]
uses two clauses of width at most three. Fix $z_0=0$ and construct two formulas by fixing $z_h$ respectively to zero and one. Only the $h-1$ internal path positions are new variables. Thus a formula on $n$ original variables produces queries on $n+h-1$ variables, irrespective of the clauses' total width.

Fix an original assignment $x$ satisfying $B$, and let $a_0(x),a_1(x)$ be its numbers of path extensions. If all coverage clauses hold, every edge enforces equality: the zero right boundary permits exactly one extension, while the one right boundary permits none. If a clause fails, its edge is unrestricted. Flipping every path value to the right of that edge swaps the boundary condition and preserves all other edge constraints. This is a bijection, so the two extension counts are equal. Assignments violating $B$ have no extensions. In summary,
\[
 a_0(x)-a_1(x)=
 \begin{cases}
  1,&x\text{ satisfies every original condition},\\
  0,&\text{otherwise}.
 \end{cases}
\]
Summing this pointwise identity gives the desired model count. The cancellation is exact for each original assignment; it does not rely on an aggregate balance between unrelated assignments.

The same idea handles more general local witnesses. If a condition is a disjunction of terms of length at most $d$, let every term trigger equality on its edge:
\[
 (\ell_1\wedge\cdots\wedge\ell_t)
 \ \Longrightarrow\ (z_{j-1}=z_j),\qquad t\le d.
\]
The two implication clauses now have width at most $d+2$, but require no variable per term. Several true terms only repeat the same equality requirement; they do not create several extensions. Thus the number of simultaneous conditions controls the auxiliary-variable budget, while the size of a witness controls the clause width. For $d>1$ we invoke the counter at that fixed width, rather than split all clauses into $3$-CNF and risk losing the variable bound.

\Needspace{14\baselineskip}
\begin{example}[Two coverage conditions]\label{ex:path}
Take $U=\{1,2,3\}$ and $\F=\{\{1,2\},\{2,3\}\}$. Exclusion forbids choosing either adjacent pair. There is one internal path variable, and the extension counts are
\begin{center}
\begin{tabular}{lcc}
\toprule
Original assignment & $a_0(x)$ & $a_1(x)$\\
\midrule
$010$ or $101$ (each) & $1$ & $0$\\
$100$ or $001$ (each) & $1$ & $1$\\
$000$ & $2$ & $2$\\
Any assignment violating exclusion & $0$ & $0$\\
\bottomrule
\end{tabular}
\end{center}
The two queries have six and four models. Their difference counts exactly the hitting sets $\{2\}$ and $\{1,3\}$.
\end{example}

The representations use this mechanism in different ways. After row selection, an exact equation of demand at most $d$ forbids an excess through short clauses and certifies attainment through a term using at most $d$ literals. There is still one path edge per retained row. For local dimension-one tables, the preliminary step is different: enumerate the polynomially many assignments consistent with all unary and binary projections, and exclude each unwanted assignment by a signed exact-one equation. All equations from all tables are compressed together, not separately by table.

For graph problems, use the threshold bits $y_{u,j}=[|N(u)\cap S|\ge j]$ for a fixed range $1\le j\le D$. Bounded-width clauses force each bit upward whenever a corresponding $j$-element witness exists. A bounded-term predicate enforces the reverse implication. The correct threshold vector is therefore unique for every selection. Fixed finite/cofinite rules only inspect this vector, so there are constantly many variables and predicates per vertex, despite potentially many neighbor subsets.

Finally, exact counting of original objects requires a count-preserving source encoding, but decision does not. In upper domination, for instance, a valid selected set can have several private-neighbor certificates. The difference still counts satisfying selection--certificate pairs and is positive exactly when a valid set exists. All queries remain ordinary unweighted counting instances. Their answers have only linearly many bits, and they can be evaluated sequentially, preserving polynomial space. The formal argument below separates this shared counting mechanism from the individual source encodings.

\section{Preliminaries}\label{sec:prelim}

All logarithms are to base two, running times count bit operations, and polynomial space is measured in the explicit input length. We use
\begin{equation}\label{eq:rate-function}
 \eta(t):=\rate{t}.
\end{equation}
The offsets make this expression defined for nonnegative $t$; bounded input parameters will be handled directly. For a fixed positive constant $c$, $\eta(ct)=O_c(\eta(t))$, and $\eta(t)=o(t)$.

An exact-one instance explicitly declares its ground set $[r]$, so $r=O(L+1)$. Its family may repeat sets. An empty set forces count zero, and an empty family has count $2^r$; these rules also handle $r=0$. For $x\in\{0,1\}^r$, exact-one satisfaction means
\begin{equation}\label{eq:linear}
 \sum_{i\in S}x_i=1\qquad(S\in\F).
\end{equation}
These are integer equalities, equivalently rational equalities on Boolean assignments, not parity constraints.

For a formula $G$, $\satcount{G}$ counts assignments to all its declared variables, including unused ones. A $K$-CNF has clauses of width at most $K$ with ordinary disjunctive semantics. An empty clause is false. A \emph{term} is a conjunction of literals, and a DNF is a disjunction of terms. An empty term is true and an empty disjunction is false. All term lists below are explicit. Repeated literals may be removed and contradictory terms deleted. Our counting queries are ordinary unweighted formulas: no variables are existentially projected out.

Graphs are explicit and simple, and $v$ denotes their vertex count. Write $N(u)$ for the open neighborhood and $N[u]:=N(u)\cup\{u\}$. Digraphs are loopless and may have antiparallel arcs; their out-neighborhoods are denoted $N^+(u)$. Cardinality thresholds are part of the input. By contrast, a fixed width, local degree rule, or number of classes is independent of input size. Constants and polynomial degrees may depend on those fixed choices.

\begin{lemma}[Row selection; Jansen--Pieterse~\cite{JP16,JP19}]\label{lem:rows}
From an explicitly represented rational linear system $Ax=b$ in $r$ variables, one can select at most $r+1$ original equations in deterministic polynomial time with exactly the same rational, and hence Boolean, solutions. Applied to~\eqref{eq:linear}, the output retains original input sets.
\end{lemma}
\begin{proof}
Select original rows forming a basis of the augmented row space $(A\mid b)$. Every omitted row has the form $(a,b_0)=\sum_{j=1}^h\lambda_j(a_j,b_j)$ for selected rows. If $x$ satisfies the selected equations, then
\[
 a\cdot x=\sum_{j=1}^h\lambda_j(a_j\cdot x)
         =\sum_{j=1}^h\lambda_jb_j=b_0.
\]
Thus it satisfies every omitted equation. The reverse direction holds because the selected equations are a subset of the input. This does not assume consistency. The augmented dimension is $r+1$. Exact Gaussian elimination selects the rows in polynomial bit complexity; after clearing denominators, fraction-free elimination keeps intermediate bit lengths polynomial. The output is a subset of the original equations, not their row-echelon combinations, so their form and widths do not change.
\end{proof}

The next imported lemma allows arbitrarily many clauses. Its width-three case suffices for exact-one systems and all compact encodings in Section~\ref{sec:graphs}. The local degree theorem uses its full fixed-width form.

\begin{lemma}[Fixed-width exact counting; OpenAI~\cite{OAI26}]\label{lem:counter}
For every fixed $K\ge2$, a $K$-CNF on $N\ge20$ declared variables and $M$ clauses can be counted deterministically in
\begin{equation}\label{eq:counter}
 (N+M)^{O_K(1)}2^{O_K(N\log\log N/\sqrt{\log N})}
\end{equation}
bit operations and $(N+M)^{O_K(1)}$ bits of space, with no bound on $M/N$.
\end{lemma}
\begin{proof}
Apply the general-density part of Theorem~1.1 of~\cite{OAI26}. The source counts declared variables and allows shorter clauses by repeating a literal, so our width convention changes neither variables nor counts. For $N<20$, use direct enumeration.
\end{proof}

CNF encodings of cardinality constraints have been studied extensively~\cite{Sinz05}. We use the following functional encoding of unweighted cardinalities. A proof is included because its \emph{linear} variable budget is needed repeatedly; a separate sequential counter with a growing threshold would not give the same guarantee. Linear-size cardinality circuits also occur in~\cite[proof of Corollary~7]{Anth26}.

\begin{lemma}[Linear-size cardinality encoding]\label{lem:cardinality}
For $x_1,\ldots,x_v$ and an input threshold $k$, each of $\sum_i x_i\le k$, $\sum_i x_i\ge k$, and $\sum_i x_i=k$ has a uniformly constructible $3$-CNF encoding with $O(v)$ auxiliary variables and clauses. Each satisfying input assignment has exactly one auxiliary extension, and each other input has none. A fixed number of such tests still costs $O(v)$ variables.
\end{lemma}
\begin{proof}
First handle thresholds outside $[0,v]$ as constant conditions. Pad the input to a power of two using constant zeros. A balanced binary addition tree computes the sum. At level $j$, it performs $O(v/2^j)$ additions of $O(j)$-bit numbers. Ripple-carry addition uses $O(j)$ fan-in-two gates per addition, for total size
\[
 O\!\left(v\sum_{j\ge1}\frac{j}{2^j}\right)=O(v).
\]
Comparison of the resulting $O(\log(v+1))$ bits with $k$ adds $O(\log(v+1))$ gates. Use AND, OR, and XOR gates and negations. A full truth-table encoding for each gate uses a constant number of clauses of width at most three, and one variable for its output. Require the final output to be true. For any input assignment, all gate values are uniquely determined in topological order, so the extension exists precisely when the test succeeds and is then unique. Constant-size inputs are immediate. Repeat this construction for a fixed number of tests.
\end{proof}

The cardinality lemma concerns unweighted counts. It does not assert a linear-size encoding for sums with unrestricted binary weights.

\section{The two-boundary counting identity}\label{sec:identity}

The identity separates two costs: the number of conditions that must all hold, and the number of alternative witnesses for each condition. Only the first costs path variables. We prove the bounded-term version once; wide clauses are its one-literal special case.

\begin{theorem}[Two-query identity for bounded-length terms]\label{thm:boundary}
Let $B$ be a CNF of width at most $b$ on $n$ declared variables. For $j\in[h]$, let
\[
 P_j(x):=\bigvee_{T\in\terms_j}\ \bigwedge_{\ell\in T}\ell,
 \qquad H:=B\wedge\bigwedge_{j=1}^h P_j,
\]
where every term has at most $d$ literals and all term lists are explicit. Suppose $h\ge1$, and let $m_B$ be the number of clauses of $B$. There are two nonadaptive ordinary unweighted CNF queries satisfying
\begin{equation}\label{eq:general-identity}
 \satcount{H}=\satcount{Q_0}-\satcount{Q_1},
\end{equation}
with the exact variable budget and clause bounds
\begin{equation}\label{eq:general-budget}
 N=n+h-1,\qquad
 K=\max\{b,d+2\},\qquad
 M_\beta\le m_B+2\sum_{j=1}^h|\terms_j|.
\end{equation}
They are constructible in polynomial time in the explicit representation. For $h=0$, one query to $B$ suffices.
\end{theorem}
\begin{proof}
Introduce fresh variables $z_1,\ldots,z_{h-1}$. For $\beta\in\{0,1\}$ set the endpoints to constants,
\begin{equation}\label{eq:boundaries}
 z_0:=0,\qquad z_h:=\beta,
\end{equation}
and, for a term $T$, put $\overline T:=\bigvee_{\ell\in T}\neg\ell$. Define
\begin{equation}\label{eq:term-queries}
 Q_\beta:=B\wedge\bigwedge_{j=1}^h\ \bigwedge_{T\in\terms_j}
 \bigl[(\overline T\vee\neg z_{j-1}\vee z_j)
       \wedge(\overline T\vee z_{j-1}\vee\neg z_j)\bigr].
\end{equation}
Substitute the endpoint constants and delete tautological clauses, but retain all declared variables. The pair for $T$ expresses
\[
 \left(\bigwedge_{\ell\in T}\ell\right)
 \Longrightarrow z_{j-1}=z_j.
\]
It has width at most $d+2$, introduces no variable for $T$, and contributes at most two clauses. The endpoints are constants rather than variables. This proves~\eqref{eq:general-budget}.

For correctness, fix the original assignment $x$ and let $a_\beta(x)$ be its number of extensions to the internal path variables. If $B(x)=0$, both extension counts are zero. Otherwise, edge $j$ enforces equality precisely when $P_j(x)=1$: some true term activates equality, whereas if all terms are false every pair of clauses on that edge is automatically satisfied.

If all predicates are true, the whole path equals its left endpoint zero. Thus the all-zero internal assignment is the unique extension for $\beta=0$, and there is no extension for $\beta=1$.

If some predicate is false, fix one such index $j$ and map an extension under boundary $\beta$ to one under boundary $1-\beta$ by
\begin{equation}\label{eq:suffix-flip}
 z_i'=\begin{cases}
 z_i,&0\le i<j,\\
 1-z_i,&j\le i\le h.
 \end{cases}
\end{equation}
The left endpoint stays zero, and the right endpoint changes as required. On every edge other than $j$, either both endpoints are flipped or neither is, preserving equality requirements. Edge $j$ is unrestricted. The map is its own inverse and fixes the original assignment, so it is a bijection of the two extension sets. Hence $a_0(x)=a_1(x)$. This also covers $h=1$: there are no internal variables and the map changes only the fixed right boundary. Empty term lists give unrestricted edges, and an empty term forces equality, consistently with the Boolean conventions.

We have therefore proved the pointwise identity
\[
 a_0(x)-a_1(x)=\ind[H(x)=1].
\]
Summing over every assignment to the $n$ declared variables proves~\eqref{eq:general-identity}. All construction steps traverse explicit terms and clauses, so their bit complexity is polynomial in the representation length.
\end{proof}

\begin{lemma}[Wide-clause specialization]\label{lem:path}
Let $B$ be a $3$-CNF on $n$ declared variables and let $C_1,\ldots,C_h$ be arbitrary-width clauses on those variables, with either sign of literal. For $h\ge1$, two ordinary $3$-CNFs satisfy
\begin{equation}\label{eq:path-identity}
 \satcount{B\wedge C_1\wedge\cdots\wedge C_h}
 =\satcount{Q_0}-\satcount{Q_1},
\end{equation}
each with exactly $n+h-1$ declared variables and at most $m_B+2\sum_j|C_j|$ clauses.
\end{lemma}
\begin{proof}
Regard each literal of $C_j$ as a one-literal term in $P_j$ and apply Theorem~\ref{thm:boundary} with $b=3,d=1$. In this case the queries are explicitly
\begin{equation}\label{eq:queries}
 Q_\beta:=B\wedge\bigwedge_{j=1}^h\ \bigwedge_{\ell\in C_j}
 \bigl[(\neg\ell\vee\neg z_{j-1}\vee z_j)
       \wedge(\neg\ell\vee z_{j-1}\vee\neg z_j)\bigr],
\end{equation}
with the boundaries in~\eqref{eq:boundaries}.
\end{proof}

For term length greater than one, Theorem~\ref{thm:boundary} is a fixed-width reduction, \emph{not necessarily a width-three reduction}. Introducing a new variable to split every longer clause could destroy its linear variable budget. We instead apply Lemma~\ref{lem:counter} at the resulting fixed width. For a fixed source assignment, all cancellation takes place between ordinary nonnegative extension counts; neither weighted counting nor projection is used.

\subsection{Relation to marker--parity cancellation}\label{subsec:marker-relation}

The public wide-clause construction of~\cite{Ono26} introduces a marker $m_j$ that may be true only when the $j$th clause is false, and subtracts the odd-marker count from the even-marker count. For a fixed assignment with $u$ false clauses, its signed contribution is $(1-1)^u$. A sequential parity circuit uses $h$ markers and $h-1$ outputs, giving $n+2h-1$ declared variables per query.

The same construction applies to a bounded-term predicate by adding
\[
 \neg m_j\vee\overline T\qquad(T\in\terms_j).
\]
Let $z_j=m_1\oplus\cdots\oplus m_j$, with $z_0=0$ and $z_h=\beta$. The change of coordinates
\[
 m_j=z_{j-1}\oplus z_j
\]
is bijective between marker assignments of parity $\beta$ and internal path assignments. A true term forces $m_j=0$, which is exactly the equality across edge $j$ in~\eqref{eq:term-queries}. Eliminating all marker variables, and treating the fixed final parity as a constant, leaves only the $h-1$ internal path variables. This derives the path realization and its count identity without treating the two cancellation mechanisms as unrelated.

For bounded-length terms, the marker implementation has width at most $\max\{b,d+1,3\}$, whereas the path uses $\max\{b,d+2\}$. The variable saving can therefore cost one unit of clause width; neither resource budget uniformly dominates the other. The improvement from $n+2h-1$ to $n+h-1$ is an explicit variable saving, not a different asymptotic exponent under Lemma~\ref{lem:counter}. In particular, marker--parity cancellation combined with row selection already implies the rate of Theorem~\ref{thm:main}. The representations developed next specify which constraints admit the required small number of predicates and bounded term length. The bounded-term extension above is an algebraic extension of the same mechanism, not a claim that the cited catalog explicitly stated it.

\begin{proposition}[Shared algorithmic transfer]\label{prop:transfer}
Fix $b,d,C$. Suppose a problem with positive integer parameter $p$, input length $L$, and $p\le(L+1)^{O(1)}$ has a polynomial-time encoding of the form in Theorem~\ref{thm:boundary}, with
\[
 p\le n,\qquad n+h\le Cp,
\]
and total representation length polynomial in $L+p$. Assume a source solution exists exactly when the encoded formula is satisfiable. Then feasibility is decidable with at most two exact-counting queries in deterministic polynomial space and time
\begin{equation}\label{eq:common-time}
 \poly(L+p+1)\,2^{O(\eta(p))}.
\end{equation}
The constants may depend on the fixed encoding parameters. If the source count equals the encoded model count, the same algorithm computes that count exactly.
\end{proposition}
\begin{proof}
For $h\ge1$, the identity returns the model count of the encoding as the difference of two answers; for $h=0$, use one answer. The query width is fixed and $p\le N\le Cp$. By Lemma~\ref{lem:counter}, each query takes the time in~\eqref{eq:common-time}; logarithms of $N$ and $p$ are comparable for large $p$, and small $p$ is handled directly. The polynomial number of input clauses contributes only a polynomial factor. The difference is positive exactly for feasible inputs.

Each answer has at most $N+1=O(p)$ bits. Construct the two queries nonadaptively, evaluate them sequentially, and subtract their exact integer answers. Construction, storage, and arithmetic use $\poly(L+p)$ space, which is polynomial in $L+1$ by the parameter assumption. Nothing in the feasibility claim requires different source solutions to have the same number of certificates; it requires only that every solution have at least one and that invalid selections have none.
\end{proof}

For witness recovery, additionally assume a polynomial-time map that decodes every satisfying assignment of the source encoding into a valid source solution. After a YES answer, fix its $n$ base variables in turn, trying zero and retaining that value precisely when a satisfying extension still exists. Additional unit clauses go in $B$ and do not alter the variable or predicate budget. After at most $n$ further tests, decode the resulting satisfying assignment. All encodings below expose the original selection or class bits, so they have the required map; one may stop as soon as these bits have been fixed. Polynomially many calls preserve~\eqref{eq:common-time}.

\Needspace{11\baselineskip}
\section{Exact-One Hitting Set and bounded-demand equations}\label{sec:exactone}\label{sec:algorithm}

We now apply the path identity to exact-one systems. Row selection bounds the number of remaining coverage clauses; pairwise exclusions handle the other half of each exact-one condition. We first prove the reduction without assuming any particular counting algorithm, and then derive the running time.

\begin{proof}[Proof of Theorem~\ref{thm:reduction}]
If $\F$ contains an empty set, return zero. If $\F$ is empty, return $2^r$. In all other cases apply Lemma~\ref{lem:rows} to~\eqref{eq:linear}, retaining original sets $S_1,\ldots,S_h$ with $1\le h\le r+1$. Their exact-one constraints have exactly the same Boolean solutions as the input.

Let $E$ be the set of unordered pairs of distinct elements that occur together in at least one retained set. Define
\begin{equation}\label{eq:exclusion}
 B(x):=\bigwedge_{\{u,v\}\in E}(\neg x_u\vee\neg x_v),
 \qquad C_j(x):=\bigvee_{i\in S_j}x_i\quad(j\in[h]).
\end{equation}
If an assignment satisfies the selected exact-one constraints, it satisfies every pairwise exclusion and every $C_j$. Conversely, $B$ allows at most one chosen element in each $S_j$, because any two distinct elements of $S_j$ form a pair in $E$; $C_j$ requires at least one. Hence
\begin{equation}\label{eq:split}
 Z(U,\F)=\satcount{B\wedge C_1\wedge\cdots\wedge C_h},
\end{equation}
where the formula retains all $r$ original variable declarations. Lemma~\ref{lem:path} now gives the two queries and the identity~\eqref{eq:main-identity}.

The resource bounds follow separately. There are $h-1$ internal path variables, so each query has
\begin{equation}\label{eq:variable-budget}
 N=r+h-1\le2r.
\end{equation}
Since $E$ contains each unordered pair at most once, $|E|\le\binom r2$. Also $\sum_j|S_j|\le hr\le r(r+1)$. The clause bound from Lemma~\ref{lem:path} is therefore
\begin{equation}\label{eq:clause-budget}
 M_b\le |E|+2\sum_{j=1}^h|S_j|
       \le\binom r2+2r(r+1)=O(r^2),\qquad b\in\{0,1\}.
\end{equation}
Literal incidences contribute clauses but no extra variables. Original variables that are absent from the selected sets, or disappear during boundary simplification, remain declared. Thus their choices retain the correct multiplicity without a separate correction factor.

Row selection has polynomial bit complexity. The pairs can be enumerated with $O(r^3)$ pair operations and deduplicated deterministically; constructing both query lists is also polynomial-time work. Neither construction depends on an oracle answer. Finally, each answer is at most $2^N\le2^{2r}$ and therefore has at most $2r+1$ bits. Subtracting the two exact integers takes polynomial time and yields $Z(U,\F)$, a nonnegative integer at most $2^r$.
\end{proof}

Algorithm~\ref{alg:count} records the complete procedure. The subroutine $\textsc{Count3CNF}$ returns an exact model count, not a satisfiability answer.

\begin{algorithm}[H]
\caption{Exact counting for Exact-One Hitting Set}\label{alg:count}
\begin{algorithmic}[1]
\Require Explicit ground set $U=[r]$ and family $\F$
\Ensure $Z(U,\F)$
\If{$\emptyset\in\F$}
  \State \Return $0$
\EndIf
\If{$\F=\emptyset$}
  \State \Return $2^r$
\EndIf
\State Select original sets $S_1,\ldots,S_h$ using Lemma~\ref{lem:rows}.
\State Form $B$ and $C_1,\ldots,C_h$ as in~\eqref{eq:exclusion}.
\State Construct $Q_0,Q_1$ using~\eqref{eq:queries}, retaining all declared variables.
\State $A_0\gets\textsc{Count3CNF}(Q_0)$.
\State $A_1\gets\textsc{Count3CNF}(Q_1)$, after the first call has finished.
\State \Return $A_0-A_1$
\end{algorithmic}
\end{algorithm}

\begin{proof}[Proof of Theorem~\ref{thm:main}]
Theorem~\ref{thm:reduction} proves correctness: Algorithm~\ref{alg:count} returns $Z(U,\F)$, whose positivity decides feasibility. After the trivial cases, its source encoding has $n=r$ variables, $1\le h\le r+1$ coverage clauses, and polynomial representation length. Thus $r\le n$, $n+h\le3r$, and the hypotheses of Proposition~\ref{prop:transfer} hold with fixed widths. That proposition gives~\eqref{eq:main-time} and polynomial space; the factor polynomial in $L+r$ is polynomial in $L+1$ because the ground set is explicit. In particular, the two query variables satisfy $r\le N\le2r$ and their clause counts are $O(r^2)$, so the imported exponent is $O(\eta(r))$. The encoding preserves the full solution set, not merely existence, and therefore the transfer gives exact counting. Bounded $r$ may be handled directly.
\end{proof}

\subsection{Signed exact satisfiability}\label{subsec:signed}

A signed $\XSAT$ instance is a family of clauses of positive and negative literals, with the requirement that exactly one literal in each clause be true. We assume that distinct literals in a clause have distinct underlying variables. All $r$ variables are explicitly declared, and the number and widths of the clauses are unrestricted.

\Needspace{10\baselineskip}
\begin{theorem}[Signed $\XSAT$]\label{thm:signed}
The count of a signed $\XSAT$ instance can be computed deterministically in polynomial space and time~\eqref{eq:main-time}. Except for the empty-family and empty-clause cases, it is the difference of two $3$-CNF counts, each on at most $2r$ variables and with $O(r^2)$ clauses.
\end{theorem}
\begin{proof}
The empty-family and empty-clause cases give $2^r$ and zero, respectively. Consider a nonempty clause whose positive and negative literals use the disjoint variable sets $P$ and $D$. Its exact-one requirement is
\begin{equation}\label{eq:signed-row}
 \sum_{i\in P}x_i+\sum_{i\in D}(1-x_i)=1,
 \qquad\text{equivalently}\qquad
 \sum_{i\in P}x_i-\sum_{i\in D}x_i=1-|D|.
\end{equation}
The coefficients belong to $\{-1,0,1\}$ and the right-hand side has magnitude at most $r$. Lemma~\ref{lem:rows} therefore selects $h\le r+1$ original clauses in polynomial bit complexity, preserving all Boolean solutions.

For each pair of literals $\ell,\ell'$ in a retained clause, put $(\neg\ell\vee\neg\ell')$ into $B$, removing duplicates. For each unordered pair of underlying variables there are four possible sign combinations, so $B$ has at most $4\binom r2$ clauses. Let $C_j$ be the ordinary disjunction of the literals in retained clause $j$. An assignment satisfies $B\wedge\bigwedge_j C_j$ exactly when it makes one literal true in each retained clause, hence exactly when it solves the original signed instance.

Lemma~\ref{lem:path} applies without a change: it depends on a literal's truth value, not on its sign. It produces two queries with
\[
 N=r+h-1\le2r,\qquad
 M_b\le4\binom r2+2hr=O(r^2),
\]
whose count difference is the desired answer. These are the same asymptotic resource bounds used in the proof of Theorem~\ref{thm:main}, so the same time and space analysis applies.
\end{proof}

The distinct-variable convention is part of this statement; repeated occurrences of a variable within one exact-one clause are not treated as independent choices.

\begin{corollary}[Binary side constraints]\label{cor:binary-side}
The signed exact-one count may additionally be restricted by any explicitly given $2$-CNF on the same $r$ variables. After direct treatment of empty cases, at most two $3$-CNF queries suffice, each on at most $2r$ variables and with $O(r^2)$ clauses. The time and space bounds of Theorem~\ref{thm:main} are unchanged.
\end{corollary}
\begin{proof}
Keep the side clauses in $B$ and apply row selection only to the exact-one equations. The argument above preserves their intersection with $B$. After simplification and deduplication there are only $O(r^2)$ possible unary and binary clauses, so the clause budget is unchanged. If no equations remain, count the side formula directly. A contradictory side clause gives count zero.
\end{proof}

\subsection{Exact Cover measured by candidate sets}\label{subsec:cover}

Let $V$ be a universe and $A_1,\ldots,A_s\subseteq V$ an explicitly listed, indexed family of candidate sets. An exact cover is a set of indices $I\subseteq[s]$ such that each $v\in V$ lies in exactly one $A_i$ with $i\in I$. Sets with different indices are distinct choices even when their contents agree. The parameter here is the number $s$ of candidate sets, not $|V|$.

\begin{corollary}[Exact Cover]\label{cor:cover}
For an Exact Cover instance of input length $L$ with $s$ indexed candidate sets, the number of exact covers can be computed deterministically in polynomial space and time
\begin{equation}\label{eq:cover-time}
 \poly(L+1)\,2^{O\!\left(\rate{s}\right)}.
\end{equation}
The decision problem has the same bound, without a restriction on the universe size or the number of incidences.
\end{corollary}
\begin{proof}
Take $[s]$ as the selectable ground set and form, for every $v\in V$, the scope
\[
 S_v:=\{i\in[s]:v\in A_i\}.
\]
A set of indices $I$ hits each $S_v$ exactly once if and only if the corresponding candidate sets cover each $v$ exactly once. The correspondence uses the same index set $I$ in both directions, so it preserves the count bijectively. Transposing the incidence list takes polynomial time and gives an explicitly represented instance of polynomial length. Apply Theorem~\ref{thm:main} with $r=s$.

An uncovered universe element creates an empty scope and correctly forces count zero. An empty candidate set, when allowed, creates an unconstrained declared variable; it is counted as an optional indexed choice, as required by the stated definition.
\end{proof}

\subsection{Fixed, row-dependent exact demands}\label{subsec:demands}

Generalized exact satisfiability allows different clauses to require different numbers of true literals~\cite{HS21}. We also allow positive integer coefficients supplied in binary. The equations have the form
\begin{equation}\label{eq:demand-input}
 \sum_{i\in S_j}a_{ji}\ell_{ji}=b_j,
 \qquad a_{ji}\in\mathbb Z_{>0},
\end{equation}
where literals are evaluated as zero or one. The number of rows and their widths are unrestricted. The fixed-demand assumption is imposed \emph{after} the following solution-preserving normalization.

Within a row, aggregate the coefficients of $x_i$ and $\neg x_i$ as $p_i$ and $q_i$. Their contribution is
\[
 p_i x_i+q_i(1-x_i)
 =\min\{p_i,q_i\}+|p_i-q_i|\ell_i,
\]
where $\ell_i=x_i$ if $p_i>q_i$ and $\ell_i=\neg x_i$ otherwise; omit a zero difference. Subtract the constants from the target. If the resulting target is outside $[0,A_j]$, where $A_j$ is the sum of the remaining positive coefficients, return zero. Complementing every remaining literal changes the target to $A_j-b_j$, so use whichever target is smaller. Constant true rows may be deleted. The resulting rows use distinct underlying variables. All this takes polynomial bit time even when the original coefficients are large.

\begin{theorem}[Fixed-demand exact counting]\label{thm:fixed-demand}
Fix an integer $d\ge1$. For a system~\eqref{eq:demand-input} on $r$ declared variables whose normalized row demands are at most $d$, its exact solution count is recoverable by at most two ordinary CNF counting queries, each on at most $2r$ variables, of width at most $\max\{3,d+2\}$, and with $O_d((r+1)^{d+2})$ clauses. The construction takes polynomial time for fixed $d$, and the resulting algorithm uses polynomial space and time
\[
 \poly_d(L+1)\,2^{O_d(\eta(r))}.
\]
The demands may differ among rows, and the original positive coefficients are binary encoded. Only $d$ is fixed independently of the input.
\end{theorem}
\begin{proof}
Perform the normalization above. Return $2^r$ if no rows remain and handle $r=0$ directly. In each row replace any coefficient larger than its target $b_j$ by $b_j+1$. This preserves the Boolean solutions of that row: a literal with such a coefficient must be false in either system. Apply Lemma~\ref{lem:rows} to the \emph{capped, normalized} affine equations, retaining $h\le r+1$ of those rows. This step is after capping; the retained rows have exactly the same Boolean solutions as the full normalized system. Denote a retained row's contribution by $s_j(x)=\sum_i a_{ji}\ell_{ji}$.

To enforce $s_j(x)\le b_j$, put into a CNF $B$ the clauses
\begin{equation}\label{eq:demand-upper}
 \bigvee_{i\in A}\neg\ell_{ji}
 \quad\left(A\subseteq S_j,\ |A|\le b_j+1,
                  \sum_{i\in A}a_{ji}>b_j\right).
\end{equation}
These clauses never reject a contribution at most $b_j$. Conversely, suppose the true literals contribute more than $b_j$ and choose an inclusion-minimal overweight subset $A$. Removing any one item leaves total weight at most $b_j$. Positivity and integrality of the weights then imply $|A|-1\le b_j$, so its forbidden clause occurs in~\eqref{eq:demand-upper}. Thus these clauses enforce exactly the upper bound, and have width at most $d+1$.

For attainment of the target, require
\begin{equation}\label{eq:demand-attainment}
 P_j(x):=
 \bigvee_{\substack{A\subseteq S_j,\ |A|\le d\\
                    \sum_{i\in A}a_{ji}=b_j}}
       \ \bigwedge_{i\in A}\ell_{ji}.
\end{equation}
When $b_j=0$, the empty subset supplies the empty, true term. Under the upper bound, $P_j$ is equivalent to $s_j(x)=b_j$: a true term already contributes $b_j$; conversely, if equality holds, the entire set of true literals is such a term because each positive weight is at least one and hence at most $b_j\le d$ literals are true. If no subset attains the target, $P_j$ is false, as required. Every term has length at most $d$.

The formula $B\wedge\bigwedge_{j=1}^h P_j$ is therefore pointwise equivalent to the original system on its $r$ declared variables, without certificates. For fixed $d$, there are $O_d((r+1)^{d+2})$ upper clauses and $O_d((r+1)^{d+1})$ terms. Theorem~\ref{thm:boundary} gives $N=r+h-1\le2r$ and the stated fixed width, and Proposition~\ref{prop:transfer} gives the time, count, and space guarantees. In a trivial no-row case no query is needed.
\end{proof}

An instance can thus mix demands zero, one, and two, rather than requiring a uniform demand. This is a fixed-bound theorem, not a uniform guarantee for $d$ growing with $r$. Appendix~\ref{app:one-query} gives a different, parsimonious implementation with one query of width $\max\{3,d+1\}$ and $O_d(r)$ variables, using the arithmetic lemma there. It has a lower width but not the explicit $2r$ variable budget; the two implementations offer different guarantees.

\section{Local relations of VC dimension at most one}\label{sec:vc}

An allowed table $R\subseteq\{0,1\}^w$ explicitly lists the permitted patterns on an ordered scope of $w$ distinct variables. Different tables may have arbitrary, overlapping scopes. The table has VC dimension at most one if no pair of coordinates realizes all four patterns. Empty tables force count zero; the nonempty table on a zero-length scope is tautological. The restriction is on each local table, not on the global set of solutions.

Dimension-one classes admit familiar structural descriptions~\cite{BenDavid15}. The interpolation bound $\operatorname{intdeg}(R)\le\operatorname{VCdim}(R)$ also gives affine separation on a dimension-one domain~\cite[Theorem~2.2 and Proposition~6.1]{MR20}. We prove a more specific representation for use with the preceding exact-one reduction: the nonbinary conditions have literal coefficients and right-hand side one. The proof includes the construction and its size, rather than assuming that a suitable affine description is supplied.

\subsection{Binary closure and effective enumeration}

For a nonempty table $R$, let $B_R$ forbid every missing unary or binary pattern. For instance, if coordinates $i,j$ never take the values $a,b$ together, include the clause $(x_i\ne a)\vee(x_j\ne b)$. Let
\[
 \mathcal U_R:=\{x\in\{0,1\}^w:B_R(x)=1\}.
\]
Then $R\subseteq\mathcal U_R$. The closure also has VC dimension at most one: for every coordinate pair, a pattern missing from $R$ is explicitly forbidden by $B_R$.

\begin{lemma}[Size and enumeration of the binary closure]\label{lem:vc-closure}
A Boolean class on $w$ coordinates that shatters no pair has at most $w+1$ members. Consequently, $\mathcal U_R$ has at most $w+1$ assignments and can be listed in deterministic polynomial time in the explicit description of $R$ and its scope.
\end{lemma}
\begin{proof}
For the size bound, use induction on $w$, with at most one empty pattern when $w=0$. Split a class by its last bit, obtaining prefix sets $A_0,A_1\subseteq\{0,1\}^{w-1}$. Their union still shatters no pair, so $|A_0\cup A_1|\le w$. Their intersection contains at most one prefix: two distinct common prefixes differ in some coordinate, and extending both by both last-bit values would shatter that coordinate together with the last one. Thus
\[
 |A_0|+|A_1|=|A_0\cup A_1|+|A_0\cap A_1|\le w+1.
\]

Construct $B_R$ by scanning the explicitly listed projections. Enumerate its satisfying assignments by fixing successive coordinates and using a $2$-SAT feasibility test to retain only extendible prefixes. At any depth there are at most $|\mathcal U_R|\le w+1$ retained prefixes, since different prefixes have disjoint sets of completions. There are $w$ depths and at most two tests per retained prefix. Each test concerns a polynomial-size $2$-CNF with additional unit clauses, so total time and space are polynomial. The construction keeps every scope variable, including coordinates fixed by unary clauses.
\end{proof}

The binary closure need not equal the table. For $R=\{100,010,001\}$, pairwise exclusions also allow $000$. The missing condition is exactly $x_1+x_2+x_3=1$. The next lemma shows that this form of repair always suffices.

\subsection{Separation by signed exact-one equations}

\begin{lemma}[Exact-one separation]\label{lem:vc-separation}
For each $y\in\mathcal U_R\setminus R$, one can find in polynomial time a subset $S\subseteq[w]$, with $|S|\ge3$, such that every $x\in R$ satisfies
\begin{equation}\label{eq:vc-separator}
 \sum_{i\in S}\ell_i(x)=1,
 \qquad
 \ell_i(x):=\begin{cases}x_i,&y_i=0,\\1-x_i,&y_i=1,\end{cases}
\end{equation}
whereas $y$ does not satisfy this equation.
\end{lemma}
\begin{proof}
Start with all coordinates and delete a coordinate whenever no row of $R$ agrees with $y$ on the remaining coordinates. Stop at an inclusion-minimal set $S$ on which $y$ has no agreeing row. Projection tests are scans of $R$, so this takes polynomial time. Since $y$ satisfies every unary and binary restriction, no set of size at most two has this property. Thus $|S|\ge3$.

For every $i\in S$, minimality supplies a row $x^{(i)}\in R$ agreeing with $y$ on $S\setminus\{i\}$. It must differ at $i$, since no row agrees on all of $S$. On the coordinates $S$, the disagreement vector $(\ell_j(x^{(i)}))_{j\in S}$ therefore has its sole one at $i$.

Every row of $R$ disagrees with $y$ somewhere on $S$. Suppose one row disagreed at two distinct coordinates $i,j\in S$. Its pair of disagreement bits would be $11$, while $x^{(i)}$ and $x^{(j)}$ supply $10$ and $01$. A third coordinate $k\in S\setminus\{i,j\}$ exists, and $x^{(k)}$ supplies $00$. Complementing coordinates does not change whether they are shattered. These four patterns contradict the dimension assumption on $R$. Thus every row disagrees in exactly one position of $S$, proving~\eqref{eq:vc-separator}. The assignment $y$ has zero disagreements and is excluded.
\end{proof}

\begin{theorem}[Binary clauses and exact-one representation]\label{thm:vc-representation}
Every nonempty explicitly listed Boolean relation $R\subseteq\{0,1\}^w$ of VC dimension at most one has a polynomial-time constructible representation, on its original variables, as a conjunction of $O(w^2+w)$ unary or binary clauses and at most $w+1$ signed exact-one equations. Its Boolean solution set is exactly $R$.
\end{theorem}
\begin{proof}
Use $B_R$ and enumerate $\mathcal U_R$ by Lemma~\ref{lem:vc-closure}. For each $y\in\mathcal U_R\setminus R$, add the equation in Lemma~\ref{lem:vc-separation}. Every row of $R$ satisfies all these conditions. An assignment outside $\mathcal U_R$ fails $B_R$; an assignment in $\mathcal U_R\setminus R$ fails its own separating equation. Hence the solution set is exactly $R$. The number of equations is at most $|\mathcal U_R|\le w+1$, and each has width at most $w$. All stages are polynomial in the explicit input length. The zero-coordinate nonempty relation needs no conditions.
\end{proof}

This syntax does not assert that every conjunction of binary clauses and exact-one equations has VC dimension one. It is a representation in the needed direction. In particular, affine interpolation alone does not specify the literal coefficients, right-hand side, or clause count obtained here.

\subsection{Many overlapping local constraints}

\begin{proof}[Proof of Theorem~\ref{thm:vc-intro}]
Reject an empty allowed table and discard zero-scope tautologies. Apply Theorem~\ref{thm:vc-representation} to each remaining table on its own scope, retaining every global variable declaration. Collect all binary clauses in one formula $B_0$, and all signed exact-one equations in one system. This representation is pointwise equivalent to the original conjunction, regardless of scope overlaps. Its length is polynomial in $L$, since every table and scope is explicit.

Deduplicate the clauses in $B_0$; there are only $O(r^2)$ distinct unary and binary clauses on $r$ variables. Apply row selection to the union of \emph{all} separating equations, not independently to each table. At most $r+1$ equations remain, each still a signed exact-one equation on distinct variables. This preserves all solutions of those equations and therefore also their intersection with $B_0$. Corollary~\ref{cor:binary-side} gives the two queries, each with at most $2r$ variables and $O(r^2)$ clauses. If no separating equation remains, $B_0$ itself is one query; tautological or inconsistent cases are handled directly. The imported counter and the shared transfer give the asserted exact count, bit time, and polynomial space.
\end{proof}

Local dimension one does not imply few global assignments. For example, the singleton table $\{0,1\}$ on each of $r$ disjoint scopes permits all $2^r$ assignments. The gain comes from the restricted \emph{form} of each local relation and global row selection, not from enumerating the global solution set. General dimension-one structure and affine interpolation retain their existing attribution; the explicit syntax and its query budget are the representation guarantee proved in this section.

\section{Fixed local degree constraints}\label{sec:local}

Fixed finite/cofinite neighbor-count rules distinguish only finitely many small counts and one tail interval, as in the established vertex-subset and partition frameworks~\cite{BXTV13,FMMNSSW25}. We encode a count by threshold bits rather than one-hot states. The invariant is direct: bit $j$ is true exactly when the count is at least $j$. There are $D$ such bits per vertex for a fixed cutoff $D$, with no variable for an edge or neighbor subset.

\subsection{Functional threshold bits}

Let $x_u$ indicate membership in $S\subseteq V(G)$, fix $D\ge1$, and write $c_u:=|N(u)\cap S|$.

\begin{lemma}[Threshold representation of neighbor counts]\label{lem:truncated}
For a $v$-vertex graph with $v\ge1$ and fixed $D$, introduce $Dv$ variables $y_{u,j}$, $1\le j\le D$. There is a CNF $B_D$ of width at most $D+1$ together with exactly $Dv$ explicit DNF predicates of term length at most $D$ such that every assignment to $x$ has exactly one satisfying extension to $y$. That extension obeys
\begin{equation}\label{eq:threshold-semantics}
 y_{u,j}=1\quad\Longleftrightarrow\quad c_u\ge j.
\end{equation}
The total number of clauses and terms is $O_D(v^{D+1}+v)$ and the construction is polynomial-time for fixed $D$.
\end{lemma}
\begin{proof}
For each $u,j$, put into $B_D$ all clauses
\begin{equation}\label{eq:threshold-forward}
 y_{u,j}\vee\bigvee_{w\in A}\neg x_w
 \qquad\left(A\in\binom{N(u)}j\right),
\end{equation}
and require the predicate
\begin{equation}\label{eq:threshold-reverse}
 P_{u,j}:=\neg y_{u,j}\ \vee\!
 \bigvee_{A\in\binom{N(u)}j}\ \bigwedge_{w\in A}x_w.
\end{equation}
The first family forces $y_{u,j}=1$ when some $j$ selected neighbors exist. The second condition forces their existence when $y_{u,j}=1$.

More explicitly, fix $x$. If $c_u\ge j$, an all-selected $j$-subset forces $y_{u,j}=1$ in~\eqref{eq:threshold-forward}; that subset also satisfies~\eqref{eq:threshold-reverse}. If $c_u<j$, the disjunction of $j$-witnesses is false, so~\eqref{eq:threshold-reverse} forces $y_{u,j}=0$; every forward clause already has a false-selected neighbor and is satisfied. If $j>|N(u)|$, there are no forward clauses and the reverse predicate is simply $\neg y_{u,j}$, giving the same conclusion. This proves existence and uniqueness of all bits. Their monotone order follows from~\eqref{eq:threshold-semantics}; it need not be separately enforced.

Each forward clause has width $j+1\le D+1$. The literal $\neg y_{u,j}$ is a one-literal term and each other term has length $j\le D$. There are $Dv$ predicates and $O_D(v^{D+1}+v)$ enumerable clauses and terms. No witness subset receives an auxiliary variable.
\end{proof}

For a correct threshold vector, the truncated count $s=\min\{D,c_u\}$ is recognized by the following term of length at most two:
\begin{equation}\label{eq:threshold-state}
 T_{u,s}:=
 \begin{cases}
 \neg y_{u,1},&s=0,\\
 y_{u,s}\wedge\neg y_{u,s+1},&1\le s<D,\\
 y_{u,D},&s=D.
 \end{cases}
\end{equation}
Exactly one of these terms is true. The last term represents all counts at least $D$, not just count $D$.

\subsection{Vertex-subset rules and their size distributions}

Fix finite or cofinite $\sigma,\rho\subseteq\Nzero$. A $(\sigma,\rho)$-set $S$ satisfies
\begin{equation}\label{eq:sigma-rho}
 |N(u)\cap S|\in\sigma\quad(u\in S),\qquad
 |N(u)\cap S|\in\rho\quad(u\notin S).
\end{equation}
Choose a fixed cutoff $D\ge1$ beyond which membership in both sets is constant.

\begin{corollary}[Fixed $(\sigma,\rho)$-sets]\label{cor:sigmarho}
For every fixed $\sigma,\rho$, feasible sets satisfying an input upper, lower, or exact bound on $|S|$ can be counted exactly in deterministic polynomial space and time
\[
 \poly_{\sigma,\rho}(L+1)\,2^{O_{\sigma,\rho}(\eta(v))}.
\]
For $v\ge1$, each of at most two counting queries has
\begin{equation}\label{eq:sigmarho-budget}
 N=(2D+1)v+a-1,\qquad K=D+2,\qquad M=v^{O_D(1)},
\end{equation}
where $a=O(v)$ counts functional size-test variables and is zero without a test. The empty graph is handled directly. Computing the entire cardinality distribution takes polynomially many such queries within the same asymptotic time and space.
\end{corollary}
\begin{proof}
Start with Lemma~\ref{lem:truncated}. If state $s$ is disallowed by $\sigma$, forbid $x_u\wedge T_{u,s}$; if it is disallowed by $\rho$, forbid $\neg x_u\wedge T_{u,s}$. Each prohibition is one clause of width at most three, since $T_{u,s}$ has at most two literals. For $s<D$, disallowed has its ordinary membership meaning; for $s=D$, it means the entire tail is disallowed. Uniqueness of the threshold vector makes these clauses pointwise equivalent to~\eqref{eq:sigma-rho}. Add the functional size test when required.

There are $n=(D+1)v+a$ base variables and $h=Dv$ predicates. The bounded part has width at most $\max\{D+1,3\}$ and all terms have length at most $D$. Theorem~\ref{thm:boundary} gives~\eqref{eq:sigmarho-budget}; Proposition~\ref{prop:transfer} gives the running time. Threshold bits and gates are uniquely determined, so the difference counts sets, not multiple witnesses. To obtain all sizes, repeat with $|S|=k$ for $k=0,\ldots,v$. There are only $v+1$ coefficients, each with $O(v+1)$ bits, so the calls and output use the stated resources.
\end{proof}

Independent domination uses $\sigma=\{0\}$ and $\rho=\{1,2,\ldots\}$. The choice $\sigma=\{d\}$ and $\rho=\Nzero$ requires the selected set to induce a $d$-regular graph. The threshold encoding saves one state variable per vertex compared with a $(D+1)$-way one-hot encoding, while retaining functional extensions; this is a budget improvement, not a different asymptotic rate.

\subsection{Fixed-class degree-constraint matrices}

Fix $q\ge1$ and a matrix $\templ\in(2^{\Nzero})^{q\times q}$ with finite or cofinite entries. A labeled partition $V_1,\ldots,V_q$ is feasible when
\begin{equation}\label{eq:degree-template}
 u\in V_i\quad\Longrightarrow\quad
 |N(u)\cap V_j|\in\templ[i,j]\qquad(i,j\in[q]).
\end{equation}
Classes may be empty unless excluded by input bounds $\ell_i\le |V_i|\le u_i$. The matrix and $q$ are fixed; the graph and bounds vary. Choose a common constant tail cutoff $D\ge1$ for all entries.

\begin{theorem}[Fixed degree-template counting]\label{thm:matrix}
For fixed $q,\templ$, feasible labeled partitions with input class-size bounds can be counted by at most two ordinary unweighted fixed-width CNF queries. For $v\ge1$, each query has
\begin{equation}\label{eq:matrix-budget}
 N=q(2D+1)v+a-1,\qquad
 K=\max\{q,D+2,3\},\qquad M=v^{O_{q,D}(1)},
\end{equation}
where $a=O_q(v)$ is the number of functional size-test variables, or zero without tests. The resulting time is
\begin{equation}\label{eq:local-time}
 \poly_{q,\templ}(L+1)\,2^{O_{q,\templ}(\eta(v))},
\end{equation}
with deterministic computation and polynomial space. The empty graph is decided directly from the bounds.
\end{theorem}
\begin{proof}
Use one-hot class variables $x_{u,1},\ldots,x_{u,q}$. For each target class $j$, apply Lemma~\ref{lem:truncated} to its selection vector, with fresh threshold bits $y_{u,j,t}$ for $1\le t\le D$. Once all classes are chosen, these bits uniquely encode the truncated number of class-$j$ neighbors of $u$. Let $T_{u,j,s}$ be the corresponding term from~\eqref{eq:threshold-state}. For every state $s$ disallowed by $\templ[i,j]$, forbid $x_{u,i}\wedge T_{u,j,s}$ with a clause of width at most three. As before, the tail is either entirely allowed or entirely disallowed. These clauses impose exactly~\eqref{eq:degree-template}. Add at most $2q$ functional class-size tests.

The bounded formula contains class constraints, forward threshold clauses, allowed-state clauses, and size circuits. Its width is at most $\max\{q,D+1,3\}$. There are
\[
 n=qv+qDv+a=q(D+1)v+a,\qquad h=qDv
\]
base variables and reverse predicates; all terms have length at most $D$. Theorem~\ref{thm:boundary} gives~\eqref{eq:matrix-budget}. Enumerating neighbor subsets is polynomial for fixed $q,D$. The encoding is functional for every feasible labeled partition, proving exact count preservation; Proposition~\ref{prop:transfer} gives time and space. Inconsistent size bounds are rejected, and other bounds are intersected with $[0,v]$.
\end{proof}

This counts labeled partitions, not partitions modulo permutation of class names. For fixed $q$, repeating over all size tuples still uses only polynomially many queries. Perfect Matching Cut is the two-class template with diagonal entries $\Nzero$ and off-diagonal entries $\{1\}$ and both classes required nonempty. Minimum or maximum feasible size follows by threshold tests even when the feasible sets themselves are not inclusion-monotone.

\subsection{A smaller encoding for regular induced subgraphs}

The general representation explains the scope; individual rules can avoid threshold bits. Fixed-degree regular induced subgraphs provide an example studied in the exact-algorithm literature~\cite{GRS12}.

\begin{proposition}[Maximum fixed-$d$ regular induced subgraph]\label{prop:regular}
For each fixed $d\ge1$, deciding whether at least $k$ vertices induce a $d$-regular graph reduces to at most two width-$(d+2)$ queries. For $v\ge1$, each has
\[
 N=2v+a-1,\qquad M=O_d(v^{d+2}),
\]
where $a=O(v)$ encodes the input size bound. Exact counting with an upper, lower, or exact size bound has the same budgets. The algorithm uses time $\poly_d(L+1)2^{O_d(\eta(v))}$ and polynomial space.
\end{proposition}
\begin{proof}
Use selection bits $x_u$. Let $B$ contain the functional size circuit and all clauses
\[
 \neg x_u\vee\bigvee_{w\in A}\neg x_w
 \qquad\left(u\in V,\ A\in\binom{N(u)}{d+1}\right).
\]
They allow at most $d$ selected neighbors at a selected vertex. Require, for each vertex,
\begin{equation}\label{eq:regular-lower}
 P_u:=\neg x_u\ \vee\!
 \bigvee_{A\in\binom{N(u)}d}\ \bigwedge_{w\in A}x_w.
\end{equation}
This allows an unselected vertex and otherwise demands at least $d$ selected neighbors. The conditions are therefore equivalent to induced $d$-regularity. There are $n=v+a$ variables and $h=v$ predicates, bounded width $d+2$, and term length at most $d$. There are $O_d(v^{d+2})$ upper clauses and $O_d(v^{d+1})$ terms. Apply Theorem~\ref{thm:boundary} and Proposition~\ref{prop:transfer}. Only functional size gates are added, so original selections are counted exactly. Empty graphs and trivial thresholds are treated directly.
\end{proof}

No connectivity condition is imposed: $d=2$ permits a disjoint union of cycles, not necessarily one Hamiltonian cycle. The constants $d,D,q$ are independent of input size. Unit clauses on selection or class bits permit witness recovery without changing the budgets. Polynomially many threshold or self-reduction calls preserve the displayed times.

\section{Representative graph applications}\label{sec:graphs}

Induced matching and perfect matching cuts show how direct formulas avoid the threshold variables of the general degree template. Upper domination instead illustrates why a decision encoding may have nonunique certificates. Shorter encodings for Independent Dominating Set, Dominating Induced Matching, and Kernel in a Digraph are retained in Appendix~\ref{app:graphs}. Existing models and the public cancellation constructions are attributed at their point of use.

\begin{corollary}[Graph decision algorithms]\label{cor:graph-algorithms}
The three problems treated in this section and the three additional problems in Appendix~\ref{app:graphs} can be decided on arbitrary explicitly represented $v$-vertex graphs in deterministic polynomial space and time
\begin{equation}\label{eq:graph-time}
 \poly(L+1)\,2^{O(\eta(v))}.
\end{equation}
Every feasibility test uses at most two ordinary $3$-CNF counting queries with $O(v)$ variables. Input size thresholds need not be fixed. A witness and an optimum threshold, when applicable, can be recovered within the same asymptotic bound.
\end{corollary}

All graphs are simple and undirected except for the kernel application. Budgets below concern constructed queries for $v\ge1$; empty inputs and trivial thresholds are handled directly. The symbol $a=O(v)$ is the actual number of functional gate variables in the indicated size test. The displayed budgets are not new characterizations of the underlying graph problems.

\subsection{Induced matching and exact size information}

An \emph{induced matching} is a matching whose endpoints induce exactly its matching edges. Maximum Induced Matching asks whether there are at least $k$ such edges. Its vertex-based representation is established in the formulation work of Ahat, Ekim, and Ta\c{s}k\i n~\cite{AET18}; exact algorithms are studied, for example, in~\cite{XT17}. The relevant resource advantage is that endpoint variables, unlike edge variables, remain linear in $v$ on dense graphs.

\begin{proposition}[Maximum Induced Matching]\label{prop:mim}
Given a graph $G$ and integer $k$, deciding whether $G$ has an induced matching of at least $k$ edges reduces to at most two ordinary $3$-CNF counting queries, each with $N=2v+a-1$ variables and $O(v^3)$ clauses.
\end{proposition}
\begin{proof}
A set $S$ is the endpoint set of an induced matching exactly when $G[S]$ is $1$-regular: every component is then a single edge, with no further edges between endpoints. This correspondence is bijective, and the matching has $|S|/2$ edges. Apply Proposition~\ref{prop:regular} with $d=1$ and $|S|\ge2k$. Its upper-degree clauses have width three, and its $v$ lower predicates are the wide clauses $\neg x_u\vee\bigvee_{w\in N(u)}x_w$. Thus the query width is three and the stated budgets follow. The size of every feasible endpoint set is automatically even. If $k\le0$, the empty matching suffices; if $k>\lfloor v/2\rfloor$, return NO.
\end{proof}

\begin{corollary}[Induced-matching size distribution]\label{cor:mim-distribution}
For an arbitrary $v$-vertex graph, let $m_k$ be its number of induced matchings with exactly $k$ edges. The full list $(m_0,\ldots,m_{\lfloor v/2\rfloor})$ is computable deterministically in polynomial space and time $\poly(L+1)2^{O(\eta(v))}$. Counting all induced matchings without a size test needs two $3$-CNF queries with $2v-1$ variables for $v\ge1$; a fixed size uses $2v+a-1$ variables with $a=O(v)$.
\end{corollary}
\begin{proof}
A $1$-regular endpoint set uniquely determines its induced matching, including the empty set and empty matching. Apply Proposition~\ref{prop:regular} with $d=1$, using no size test for the total count or the functional test $|S|=2k$ for $m_k$. Each endpoint set then has exactly one gate extension. Evaluate the $O(v+1)$ pairs of queries sequentially. Each coefficient is at most $2^v$ because endpoint sets are distinct; their bit lengths and number are polynomial. The zero-vertex graph has only the empty matching. The largest nonzero index and its coefficient give the optimum size and number of optimum solutions.
\end{proof}

This is a count-preserving refinement of the existing endpoint representation, not a separate graph characterization. Marker--parity elimination~\cite{Ono26} also gives the same exponent form when combined with that representation; the path implementation gives the stated smaller variable budget.

\subsection{Perfect matching cuts}

A \emph{perfect matching cut} is a partition $V=A\dot\cup B$ into two nonempty parts such that the set of all crossing edges is a perfect matching. Equivalently, each vertex has exactly one neighbor on the opposite side~\cite{LT22}. It is not merely a perfect matching that contains a cut. Exact algorithms for this problem include~\cite{LT22,MM23}. It is a fixed degree-template problem, but one side bit per vertex gives a smaller direct encoding.

\begin{proposition}[Perfect Matching Cut]\label{prop:pmc}
Deciding whether a graph has a partition into two nonempty parts whose crossing edges form a perfect matching reduces to at most two ordinary $3$-CNF counting queries, each with $N=3v-1$ variables and $O(v^3)$ clauses.
\end{proposition}
\begin{proof}
Use $x_u\in\{0,1\}$ to denote the side of vertex $u$. For every unordered pair of distinct neighbors $a,b\in N(u)$, put
\[
 (x_u\vee\neg x_a\vee\neg x_b)
 \wedge(\neg x_u\vee x_a\vee x_b)
\]
into $B$. If $x_u=0$, the first clause forbids two $1$-neighbors; if $x_u=1$, the second forbids two $0$-neighbors. Thus $B$ imposes the at-most-one condition. Add the $2v$ wide clauses
\begin{equation}\label{eq:pmc-wide}
 C_u^0:=x_u\vee\bigvee_{w\in N(u)}x_w,
 \qquad
 C_u^1:=\neg x_u\vee\bigvee_{w\in N(u)}\neg x_w.
\end{equation}
For either value of $x_u$, the active clause demands an opposite-side neighbor. Together, these constraints say exactly that every vertex has one such neighbor. In an undirected graph, the crossing edges therefore form a spanning $1$-regular graph, namely a perfect matching. On a nonempty graph, the same condition forces both sides nonempty. Conversely, any perfect matching cut satisfies all these clauses. An isolated vertex makes its two clauses inconsistent, as required.

The input has $n=v$ variables, $h=2v$ wide clauses, $O(v^2)$ total wide-clause length, and $O(v^3)$ clauses in $B$. Lemma~\ref{lem:path} gives $N=3v-1$ and $M=O(v^3)$; Proposition~\ref{prop:transfer} applies. The empty graph has no partition into two nonempty parts and is rejected directly.
\end{proof}

\subsection{Nonunique certificates: upper domination}

\emph{Upper Dominating Set} asks whether there is an inclusion-minimal dominating set of size at least $k$. Here minimality is by containment, not by cardinality. A selected vertex $u\in S$ has a \emph{private neighbor} $w$ if $N[w]\cap S=\{u\}$. A dominating set is inclusion-minimal exactly when every selected vertex has a private neighbor: that vertex becomes undominated if its unique selected neighbor is removed; conversely, a selected vertex with no private neighbor can be deleted without losing domination.

Exact counting and size distributions of minimal dominating sets, and hence upper domination, already appear in the public manuscript~\cite[Section on minimal dominating sets and upper domination]{PI26}. We retain the following decision encoding to illustrate nonunique certificates, not as a new qualitative algorithm. We use the $2v$-variable formulation of Burdett, Haythorpe, and Newcombe~\cite[Section~2.1]{BHN23}, expressed below as CNF conditions. Earlier work on upper domination also treats pathwidth algorithms and subexponential approximation~\cite{DLP22}; approximation and arbitrary-graph exact decision are different guarantees.

\begin{proposition}[Upper Dominating Set]\label{prop:uds}
Given a graph $G$ and integer $k$, deciding whether $G$ has an inclusion-minimal dominating set of size at least $k$ reduces to at most two ordinary $3$-CNF counting queries, each with $N=4v+a-1$ variables and $O(v^3)$ clauses. The count difference is positive exactly for a YES-instance.
\end{proposition}
\begin{proof}
Use $x_u$ for selection and $y_w$ to mark a private-neighbor witness. Let $A_{\ge k}(x,a)$ encode the size threshold. The bounded-width part is
\begin{equation}\label{eq:uds-narrow}
 B:=A_{\ge k}(x,a)\wedge
 \bigwedge_{w\in V}\ \bigwedge_{\{u,u'\}\subseteq N[w]}
 (\neg y_w\vee\neg x_u\vee\neg x_{u'}),
\end{equation}
where neighbor pairs contain distinct vertices. Add the $2v$ wide clauses
\begin{equation}\label{eq:uds-wide}
 C_w:=\bigvee_{u\in N[w]}x_u,
 \qquad
 R_u:=\neg x_u\vee\bigvee_{w\in N[u]}y_w.
\end{equation}
Suppose the encoding is satisfied, and set $S:=\{u:x_u=1\}$. The $C_w$ make $S$ dominating and the size circuit gives $|S|\ge k$. A marked vertex $w$ has at most one selected closed neighbor by~\eqref{eq:uds-narrow} and at least one by $C_w$. For every $u\in S$, clause $R_u$ supplies a marked $w\in N[u]$. Since the graph is undirected, $u\in N[w]$, so $N[w]\cap S=\{u\}$. Thus every selected vertex has a private neighbor, proving inclusion-minimality.

Conversely, for a minimal dominating set $S$ of the required size, mark one private neighbor for each selected vertex and no others. Every marked vertex has exactly one selected closed neighbor, so all clauses of $B$ hold. Domination gives the $C_w$, the chosen witnesses give the $R_u$, and the cardinality circuit has its consistent extension. This proves existence equivalence in both directions. Different markings of witnesses for the same $S$ are deliberately allowed.

There are $n=2v+a$ base variables and $h=2v$ wide clauses, with total wide-clause length $O(v^2)$. The bounded part has $O(v^3)$ clauses. Lemma~\ref{lem:path} therefore gives $N=4v+a-1$ and $M=O(v^3)$. The count difference counts valid selection--witness pairs; its positivity, not its value as a count of sets, decides the problem. Proposition~\ref{prop:transfer} supplies the time and space bound. For the empty graph, the only minimal dominating set is empty. On a nonempty graph, thresholds $k\le0$ give YES and $k>v$ give NO.
\end{proof}

In Formulation~1 of~\cite{BHN23}, substituting $y_w=1-z_w$ turns the existing private-neighbor inequalities into the conditions above: $y_w=1$ restricts the closed-neighborhood count to at most one, and $x_u\le\sum_{w\in N[u]}y_w$ requires a marked witness. The source model is therefore unchanged. The reduction counts its satisfying certificates; it does not identify all certificates belonging to the same minimal dominating set.

\begin{proof}[Proof of Corollary~\ref{cor:graph-algorithms}]
Each encoding here or in Appendix~\ref{app:graphs} has $v\le n$, $n+h=O(v)$, polynomial representation length, and a width-three part supplemented by wide clauses. Its model count is positive exactly for a feasible instance. Proposition~\ref{prop:transfer} gives the two-query, time, and space guarantees, including for the nonunique-certificate encoding.

Binary search computes each optimum size with $O(\log(v+1))$ threshold tests. For witness recovery, fix original selection or side bits by adding unit clauses to the same encoding, preserving a satisfying extension at every step. Upper-domination witness bits may remain unfixed. The final selection or class labels decode directly to a solution. At most $O(v)$ further calls are needed, preserving~\eqref{eq:graph-time}.
\end{proof}

\section{Scope and interpretation}\label{sec:discussion}

The main resource is a linear number of simultaneous conditions on a linear number of variables. Row selection supplies it for exact equations; the local-table representation recovers that equation syntax; threshold bits supply it directly for fixed local graph rules. Incidences and alternative witnesses add clauses instead of variables. Section~\ref{subsec:marker-relation} explains the exact relation to existing marker--parity cancellation~\cite{Ono26}: the smaller path representation does not by itself change the asymptotic exponent. The representations, budgets, and count-preservation properties are the concrete guarantees established here.

Exact equations, local allowed tables, threshold bits, and arithmetic gates preserve counts of original objects. Upper-domination certificates need not: only positivity is used, and stronger counting results already exist~\cite{PI26}. Neither use of the difference is a pair of satisfiability tests. Both queries can be satisfiable when the original instance is infeasible, and separate approximate counts need not approximate their difference.

Linear variable count is not subquadratic bit-size compression. Our exact-one and table queries may have $O(r^2)$ clauses and $O(r^2\log(r+2))$ bits, consistent with the generalized-kernel lower bounds of~\cite{JP16,JP19}. The dimension promise is local; the demand bound, degree rules, and class count are fixed. Appendix~\ref{app:weights} permits polynomially bounded integer weights, not unrestricted exponentially large weights. Full distributions use polynomially many queries. No graph-connectivity consequence or Exact Cover algorithm parameterized by universe size is asserted.

The same representations transfer any deterministic polynomial-space fixed-width bound $2^{g_K(N)}\poly(N+M)$ with $g_K(N)=o(N)$. For $p\le N\le Cp$ and sufficiently large $p$, $g_K(N)\le(\varepsilon/C)N\le\varepsilon p$. Thus the transfer does not depend on the particular quantitative rate of the cited counter.

\section*{AI methodology}
\phantomsection
\addcontentsline{toc}{section}{AI methodology}
Generative AI was used in research discussions, development and examination of the two-query identities, structural representations, and their applications, literature searches, proof organization, and substantive drafting and revision of the manuscript. The human authors are responsible for the final mathematical statements, originality, attribution, references, and presentation.

\clearpage
\phantomsection
\addcontentsline{toc}{section}{References}
\begingroup
\interlinepenalty=10000
\bibliographystyle{alphaurl}
\bibliography{ref}
\endgroup

\clearpage
\appendix
\section{Polynomial integer weights and a one-query variant}\label{app:weights}

This appendix records a separate arithmetic tool. It is not needed by the path proofs or the local-table representation. Its scope is polynomially bounded integer coefficients, not arbitrary binary weights. The purpose of the grouping argument is to justify a \emph{linear} auxiliary-variable budget; the individual Boolean arithmetic gates are standard.

\subsection{A linear-size weighted-sum circuit}

\begin{lemma}[Polynomial-coefficient weighted sums]\label{lem:weighted-circuit}
Fix $c>0$. Let $x_1,\ldots,x_n$ be Boolean variables, $\ell_i$ either $x_i$ or $\neg x_i$, and $0\le w_i\le n^c$ be explicitly given integers. Each comparison
\[
 \sum_{i=1}^n w_i\ell_i\ \mathrel{\bowtie}\ t,
 \qquad \bowtie\in\{\le,\ge,=\},
\]
has a uniformly constructible functional $3$-CNF encoding with $O_c(n)$ auxiliary variables and clauses. Every input assignment meeting the comparison has exactly one auxiliary extension; every other input has none. A fixed number of such comparisons still costs $O_c(n)$ variables, with the constant also depending on their number.
\end{lemma}
\begin{proof}
Handle $n\le1$ and targets outside the attainable range directly. For $n\ge2$, take
\[
 P=2^{\lceil(\log_2 n)/2\rceil},\qquad \sqrt n\le P<2\sqrt n.
\]
Expand each weight in base $P$: $w_i=\sum_a P^a d_{ia}$, $0\le d_{ia}<P$. The number of digit positions is $O_c(1)$. At each position $a$, group the literals by their common nonzero digit $q$, and compute the group population count
\[
 H_{a,q}:=\sum_{i:d_{ia}=q}\ell_i.
\]
Then
\begin{equation}\label{eq:grouped-sum}
 \sum_i w_i\ell_i=\sum_a P^a\sum_{q=1}^{P-1}qH_{a,q}.
\end{equation}

A group of $u$ literals has a population-count circuit with $O(u)$ fan-in-two gates, by the balanced addition argument of Lemma~\ref{lem:cardinality}. At a given digit position the groups are disjoint and contain at most $n$ literals in total, so all group counts cost $O_c(n)$ gates. Each $H_{a,q}$ has $O(\log n)$ bits. Multiplication by the known digit $q$ is implemented by summing at most $O(\log n)$ shifted copies, each of $O(\log n)$ bits, at cost $O(\log^2 n)$ gates. Across all positions there are at most $O_c(\sqrt n)$ nonempty groups, so this stage costs
\[
 O_c(\sqrt n\log^2 n)=O_c(n).
\]
The bound is asymptotic with a fixed constant; the finitely many smaller sizes are absorbed into that constant, not omitted from the algorithm.

Multiplication by $P^a$ changes wiring only. All nonnegative partial sums in~\eqref{eq:grouped-sum} have $O_c(\log n)$ bits because the total is at most $n^{c+1}$. Adding the $O_c(\sqrt n)$ group contributions and comparing with $t$ takes $O_c(\sqrt n\log n+\log n)=O_c(n)$ further gates. The digit lists and the circuit are constructible in polynomial bit time, including reading the target.

Use fan-in-two AND, OR, XOR, and unary NOT gates. Introduce one output variable per gate and encode its full truth table by a constant number of clauses of width at most three; pin the comparison output to true. A fixed input determines all gate values in topological order, proving the unique-extension property. A fixed number of tests uses disjoint gate variables and adds only a constant multiple of the size.
\end{proof}

\begin{corollary}[Polynomially bounded weight budgets]\label{cor:weight-budgets}
Suppose an encoding covered by Proposition~\ref{prop:transfer} has parameter $p$, $O(p)$ base variables, and $O(p)$ predicates. A fixed number of additional integer-weight tests on its base variables, with nonnegative coefficients at most $p^c$ for fixed $c$, can be imposed without changing the $\poly(L+p+1)2^{O(\eta(p))}$ time form or polynomial space. Each test instance uses at most two counting queries; all count-preservation guarantees of the source encoding remain valid.
\end{corollary}
\begin{proof}
The transfer hypothesis has $p\le n=O(p)$, so coefficients at most $p^c$ are also at most $n^c$. Use Lemma~\ref{lem:weighted-circuit} on the existing variables and put the resulting functional gate clauses into $B$. This adds $O(p)$ variables and clauses and changes its width only to at most $\max\{b,3\}$. The predicates are unchanged. Apply the shared transfer. Functional gates preserve multiplicities; they do not remove nonunique source certificates.
\end{proof}

The same argument permits a complete joint distribution for a fixed number of such weights, since each total ranges over only polynomially many integers. This uses polynomially many pairs of queries, and its output is polynomial in size; it is not a two-query guarantee for the full coefficient table. We do not use sign-weighted model counting or a randomized connectivity test in this paper. The distinction from separate monomial-interpolation and connected-domination constructions~\cite{PIConnected26} is substantive: only the ordinary, functional weight tests above are needed here, and no general best-possible running-time claim for weighted counting is made.

\subsection{One-query bounded-demand equations}\label{app:one-query}

The main proof of Theorem~\ref{thm:fixed-demand} uses a path so that its variable budget remains explicitly at most $2r$. The following alternative uses the arithmetic lemma to aggregate all deficits into a single test.

\begin{proposition}[Parsimonious bounded-demand encoding]\label{prop:one-query}
Under the hypotheses of Theorem~\ref{thm:fixed-demand}, the exact count can also be obtained from one ordinary CNF counting query of width $\max\{3,d+1\}$, with $O_d(r)$ variables and $r^{O_d(1)}$ clauses. Every satisfying assignment of the original system has exactly one auxiliary extension. The same time form and polynomial-space bound follow.
\end{proposition}
\begin{proof}
Perform the same normalization, coefficient capping, and row selection as in that theorem. Retain the upper-bound CNF~\eqref{eq:demand-upper} for the $h\le r+1$ selected rows, but omit their lower predicates. For any assignment satisfying this CNF, each deficit $b_j-s_j(x)$ is nonnegative. Hence all exact equations hold if and only if
\begin{equation}\label{eq:aggregate-deficit}
 \sum_{j=1}^h s_j(x)=\sum_{j=1}^h b_j.
\end{equation}
Indeed, this equality sets the sum of nonnegative deficits to zero.

Collect the coefficients of each original variable on the left. If its positive- and negative-literal totals are $p_i,q_i$, write the contribution as $\min\{p_i,q_i\}+|p_i-q_i|\ell_i$, and move the constant to the right. Since each capped coefficient is at most $d+1$ and each row uses a variable at most once,
\[
 |p_i-q_i|\le h(d+1)=O_d(r).
\]
Thus Lemma~\ref{lem:weighted-circuit}, with constants depending on $d$, supplies a functional $3$-CNF test of~\eqref{eq:aggregate-deficit} using $O_d(r)$ auxiliary variables. A target outside the attainable range gives count zero. Conjoining it with the upper-bound clauses yields width $\max\{3,d+1\}$ and exactly the original count. Its size is polynomial for fixed $d$. Apply the imported fixed-width counter; handle the no-row and bounded-size cases directly.
\end{proof}

For $d=2$, this alternative has width three, whereas the main path implementation has width four and an explicit $2r$ variable cap. One query and smaller width do not by themselves imply a smaller running-time constant, since the arithmetic construction has a different hidden variable constant. The two bounds should therefore be compared as stated rather than one being described as uniformly superior.

\section{Further compact graph encodings}\label{app:graphs}

These applications complete Corollary~\ref{cor:graph-algorithms}. They use the same path identity and existing local characterizations. They are collected here to keep the main text focused on structural representations rather than a list of problem-specific consequences. As in Section~\ref{sec:graphs}, $a=O(v)$ denotes the size-test gate count, and negative variable formulas are never used for the empty input.

\subsection{Independent Dominating Set}
\emph{Independent Dominating Set} asks for an independent dominating set of size at most $k$, equivalently a maximal independent set of size at most $k$. This is not the problem of finding a large independent set. Dedicated exact algorithms include~\cite{BDEP13}.

\begin{proposition}[Independent Dominating Set]\label{prop:ids}
Given a graph $G$ and integer $k$, deciding whether $G$ has an independent dominating set of size at most $k$ reduces to at most two ordinary $3$-CNF counting queries. Each constructed query has $N=2v+a-1$ variables and $O(v^2)$ clauses.
\end{proposition}
\begin{proof}
Use $x_u$ for selection and let $A_{\le k}(x,a)$ be the functional cardinality encoding. Put
\[
 B:=A_{\le k}(x,a)\wedge
       \bigwedge_{uw\in E}(\neg x_u\vee\neg x_w),
 \qquad C_u:=\bigvee_{w\in N[u]}x_w.
\]
The bounded-width part imposes independence and size; the clauses $C_u$ impose domination. Thus the encoding is satisfiable precisely for the desired sets. It has $n=v+a$ base variables and $h=v$ wide clauses, whose total length is $v+2|E|$. Lemma~\ref{lem:path} gives $N=2v+a-1$ and $M=O(v^2)$. Proposition~\ref{prop:transfer} applies. Negative thresholds give NO; for $k\ge v$, a maximal independent set witnesses YES. The empty graph is tested directly.
\end{proof}

\subsection{Dominating Induced Matching}
\emph{Dominating Induced Matching}, also called Efficient Edge Domination, requires each edge to be dominated by exactly one matching edge, counting a matching edge as dominating itself. Its characterization as a partition into an independent set and an induced $1$-regular graph is used in~\cite{XN15}.

\begin{corollary}[Dominating Induced Matching]\label{cor:dim}
Deciding whether a graph has a matching that dominates each edge exactly once reduces to at most two ordinary $3$-CNF counting queries, each with $N=2v-1$ variables and $O(v^3)$ clauses.
\end{corollary}
\begin{proof}
Take the $1$-regular selected-set encoding of Proposition~\ref{prop:regular} without a cardinality test, and add $x_u\vee x_w$ for each edge $uw$. The complement of the selected set is then independent. Each edge inside the selected set is one of its matching edges. Every other edge has exactly one selected endpoint and is incident with precisely that endpoint's matching edge. Conversely, for a dominating induced matching, its endpoints induce exactly its edges, and no edge can have both endpoints outside that set. Thus the characterization is exact. The added clauses are binary, so $n=v$, $h=v$, and $M=O(v^3)$, giving $N=2v-1$. An edgeless graph admits the empty matching, consistently with the encoding.
\end{proof}

\subsection{Directed absorption: kernels in digraphs}

A \emph{kernel} of a loopless digraph is a set $S$ with no arc between its members such that every vertex outside $S$ has an outgoing arc into $S$. We use this absorbing convention; reversing all arcs converts the opposite one. Unlike the analogous undirected condition, a kernel need not exist. Its connections with SAT, including the following source CNF, are studied in~\cite{WD12}.

\begin{proposition}[Kernel in a Digraph]\label{prop:kernel}
For a loopless digraph, deciding whether an independent set absorbs every outside vertex through an arc directed into the set reduces to at most two ordinary $3$-CNF counting queries, each with $N=2v-1$ variables and $O(v^2)$ clauses.
\end{proposition}
\begin{proof}
Use $x_u$ for membership in $S$ and form the known encoding
\[
 B:=\bigwedge_{(u,w)\in A}(\neg x_u\vee\neg x_w),
 \qquad C_u:=x_u\vee\bigvee_{w\in N^+(u)}x_w.
\]
The binary clauses say exactly that $S$ is independent. If $u\notin S$, its coverage clause requires an outgoing neighbor in $S$; if $u\in S$, the same clause is automatic. Thus satisfying assignments correspond exactly to kernels. For $v\ge1$, there are $n=v$ variables, $h=v$ wide clauses, and $O(v^2)$ total clause length. Lemma~\ref{lem:path} gives $N=2v-1$ and $M=O(v^2)$, and Proposition~\ref{prop:transfer} applies. The empty digraph has the empty kernel.
\end{proof}

\end{document}